\documentclass[aps,prx,reprint,groupedaddress,showpacs,longbibliography]{revtex4-1}

\usepackage[usenames,dvipsnames]{color}
\usepackage{graphicx,amssymb,amsmath}

\begin{document}


\title{Microwave-controlled generation of shaped single photons\\in circuit quantum electrodynamics}


\author{M. Pechal}
\email[]{mpechal@phys.ethz.ch}
\author{L. Huthmacher}
\author{C. Eichler}
\author{S. Zeytino\u{g}lu}
\author{A. A. Abdumalikov Jr.}
\author{S. Berger}
\author{A. Wallraff}
\author{S. Filipp}
\affiliation{Department of Physics, ETH Zurich, CH-8093 Zurich, Switzerland}


\date{\today}

\begin{abstract}
  Large scale quantum information processors or quantum communication networks will require reliable exchange of information between spatially separated nodes. The links connecting these nodes can be established using travelling photons which need to be absorbed at the receiving node with high efficiency. This is achievable by shaping the temporal profile of the photons and absorbing them at the receiver by time-reversing the emission process.
  Here we demonstrate a scheme for creating shaped microwave photons using a superconducting transmon-type three-level system coupled to a transmission line resonator. In a second-order process induced by a modulated microwave drive we controllably transfer a single excitation from the third level of the transmon to the resonator and shape the emitted photon. We reconstruct the density matrices of the created single photon states and show that the photons are anti-bunched. We also create multi-peaked photons with a controlled amplitude and phase. In contrast to similar existing schemes, the one presented here is solely based on microwave drives, enabling operation with fixed frequency transmons. 
\end{abstract}


\maketitle



One of the most important challenges in the rapidly developing field of quantum information processing and quantum communication is efficient quantum state transfer between spatially separated quantum bits. Even though systems incorporating only a few qubits successfully use coupling schemes based on atomic vibrational modes \cite{Blatt2008}, discrete electromagnetic modes of microwave cavities \cite{Raimond2001} or on-chip resonators \cite{Majer2007,Sillanpaa2007}, itinerant rather than localized photons are preferrable as information carriers for distributing entanglement and quantum networking over larger distances \cite{Kimble2008}.

A quantum channel between distant qubits can be established in a variety of ways. For example in heralded schemes using interference and subsequent detection of photons radiated \cite{Moehring2007,Nolleke2013} or scattered \cite{Slodicka2013} from the qubits. A deterministic approach relying on reabsorption of a photon emitted from one qubit by another has been the subject of several theoretical proposals \cite{Cirac1997,Korotkov2011a}. This scheme requires efficient generation of single photons on demand \cite{Lounis2005, Houck2007, Bozyigit2011}, their entanglement with the emitting qubit \cite{Blinov2004, Togan2010, Eichler2012b, DeGreve2012, Gao2012a, Schaibley2013} and a temporal shape of the photon which allows time-reversal of the emission process. It necessitates techniques for the generation of controllably shaped single photons which have been realized with optical photons \cite{Kuhn2002, Keller2004, Nisbet-Jones2011, Gulati2014} and shown to enable photon reabsorption \cite{Ritter2012}, albeit with a limited efficiency.


Here we utilize superconducting circuits as one of the promising platforms for quantum information processing, offering good coherence times \cite{Paik2011,Rigetti2012,Barends2013,Chang2013} and strong coupling between qubits and microwave photons \cite{Wallraff2004} in an easily controllable and compact solid-state system. In this so-called circuit quantum electrodynamics architecture \cite{Blais2004}, the field emitted from a superconducting circuit is confined to a one-dimensional transmission line without any additional need for spatial mode matching. It can easily be routed between different elements of a network and it has been recently shown that a classical microwave field can be received and stored with high fidelity \cite{Wenner2014,Inomata2014b}. For superconducting systems, photon shaping schemes have been proposed \cite{Korotkov2011a} and experimentally realized based on tunable couplers controlling the emission rate of a~photon localized in a~resonator into a~transmission line \cite{Yin2013b,Pierre2014}. Systems with a fixed resonator emission rate but a tunable coupling between the resonator and the qubit \cite{Srinivasan2011} can also be used for photon shaping. Both these approaches rely on flux tuning of a SQUID loop to achieve control over the qubit-resonator or resonator-transmission line coupling. However, because of the varying Josephson inductance of the loop, the frequency of the resonator changes along with the coupling. Therefore, to control the phase of the emitted photon as well as its envelope, the frequency shift needs to be compensated by an additional tunable parameter, such as the qubit-resonator detuning \cite{Srinivasan2013}.

In this Letter, we present an alternative, microwave-based approach to photon shaping. Both the amplitude and the phase of the emitted photon are controlled by a single phase- and amplitude-modulated microwave signal which induces an effective tunable qubit-resonator coupling via a second-order process to transfer the qubit excitation to the resonator field. Since qubit frequency and coupling remain fixed in this scheme, it can be realized in circuits without additional tuning elements.

Our device, described in more detail in Appendix \ref{appendixSys}, consists of an on-chip single-sided $\lambda/2$ transmission line resonator with resonance frequency $\omega_r/2\pi = 7.224\,\mathrm{GHz}$ and linewidth $\kappa/2\pi = 24\,\mathrm{MHz}$ coupled with strength $g/2\pi = 35\,\mathrm{MHz}$ to a transmon-type superconducting circuit \cite{Koch2007}. The transition frequency between the ground state $|g\rangle$ and the first excited state $|e\rangle$ of the transmon is tuned to $\omega_q/2\pi = 8.640\,\mathrm{GHz}$. In the presented photon shaping protocol, we make explicit use of the multi-level structure of the transmon, approximating it as an effective three-level ladder-type system with the transition frequency between the first and the second excited state $|f\rangle$ offset from $\omega_q$ by the anharmonicity $\alpha/2\pi = -421\,\mathrm{MHz}$. The direct transition between states $|g\rangle$ and $|f\rangle$ is  forbidden to first order.

\begin{figure}[b]
\includegraphics[scale=1.37]{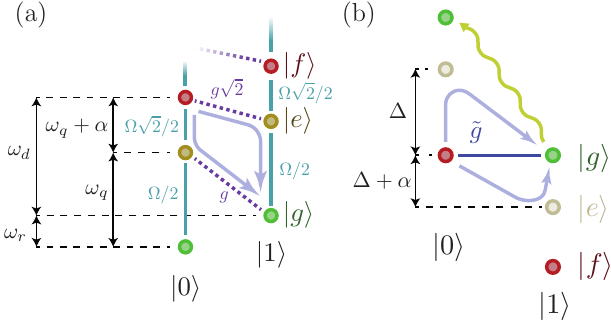}
\caption{(a) Energy level diagram of the qubit-resonator system. The matrix elements of the drive $\Omega$ and the Jaynes-Cummings coupling $g$ are indicated by solid and dotted lines, respectively, connecting the coupled bare states. The two second order paths connecting the states $|f0\rangle$ and $|g1\rangle$ are indicated by arrows. (b) In the rotating frame of the drive, the states $|f0\rangle$ and $|g1\rangle$ are nearly resonant. The second order effective coupling $\tilde{g}$ between them after adiabatic elimination of the intermediate states $|e0\rangle$ and $|e1\rangle$ is indicated by the solid blue line. The decay of the state $|g1\rangle$ into $|g0\rangle$ by photon emission is shown by the yellow arrow.}
\label{figLevelScheme}
\end{figure}

The bare eigenstates of the qubit-resonator system are coupled by the time-independent Jaynes-Cummings interaction and by an external microwave drive [Fig.~\ref{figLevelScheme}(a)] applied to the qubit through its gate line. The coupled system is in the dispersive regime \cite{Blais2004}, that is, the qubit and the resonator are detuned from each other by $\Delta = \omega_q-\omega_r \gg g$. Therefore, energy exchange between the two is strongly suppressed.

To obtain a tunable Jaynes-Cummings type coupling allowing effectively resonant swapping of excitation, we use second-order processes \cite{Gerry1990, Wu1997, Law1997}. We apply a qubit drive at the frequency $\omega_d = 2\omega_q + \alpha - \omega_r$ corresponding to the energy difference between the states $|f0\rangle$ and $|g1\rangle$, where the numbers in the kets label Fock states of the resonator. In a reference frame rotating at the frequency $\omega_d$, the Hamiltonian of the system is
\begin{align*}
  H(t) =& \delta_q b^{\dagger}b + \frac{1}{2}\alpha b^{\dagger}b^{\dagger}bb + \delta_r a^{\dagger}a +
  g(a b^{\dagger} + a^{\dagger} b) \\
  & + \frac{1}{2}(\Omega^{*}(t) b + \Omega(t) b^{\dagger}),
\end{align*}
with the effective transition frequencies of the transmon $\delta_q = \omega_q - \omega_d = -\Delta-\alpha$ and the resonator $\delta_r = \omega_r - \omega_d = -2\Delta-\alpha$. $\Omega(t) = \Omega_0(t)\mathrm{e}^{\mathrm{i}\phi(t)}$ is the complex drive strength corresponding to the real drive signal $\Omega_0(t)\cos(\omega_d t - \phi(t))$ with slowly varying amplitude $\Omega_0(t)$ and phase $\phi(t)$. The operators $a,a^{\dagger}$ are the annihilation and creation operators of the resonator and $b,b^{\dagger}$ their analogues for the transmon $b = |g\rangle\langle e| + \sqrt{2} |e\rangle\langle f|+\sqrt{3} |f\rangle\langle h|+\ldots\,$ \cite{Koch2007}. This simplified picture of the transmon as an anharmonic oscillator with a Kerr-type non-linearity is a good approximation in the limit of small anharmonicity, that is, $|\alpha|\ll \omega_q$. In the sample used for our experiment we have $|\alpha|/\omega_q\approx 1/20$ and we employ high drive amplitudes up to $\Omega/2\pi\approx 1\,\mathrm{GHz}$. To obtain good quantitative agreement between the experimental data and numerical simulations, we need to use the full model of the transmon-resonator system \cite{Koch2007}.

The states $|f0\rangle$ and $|g1\rangle$ are resonant in the rotating frame, while the intermediate states $|e0\rangle$ and $|e1\rangle$ are far off-resonant. They can therefore be adiabatically eliminated, giving rise to an effective Jaynes-Cummings type coupling between the resonator and the two-level system consisting of qubit states $|g\rangle$ and $|f\rangle$ [Fig.~\ref{figLevelScheme}(b)]. This interaction is described by the effective Hamiltonian
\begin{align}
   H_{\mathrm{eff}}(t) =& \Delta_{f0g1}(t) |f0\rangle\langle f0|\label{eqEffHam}\\
   &+ \tilde{g}(t) |f0\rangle\langle g1| + \mathrm{h.c.},\nonumber
\end{align}
which we have simplified by absorbing the parts of the AC Stark shifts independent of $\Omega$ into the renormalized frequencies of the qubit and the resonator. The remaining AC Stark shift $\Delta_{f0g1}(t)$ of the transition frequency between $|f0\rangle$ and $|g1\rangle$ is to leading order quadratic in the drive strength $\Omega$. Using perturbation theory, the effective second-order coupling $\tilde{g}(t)$ \cite{ZeytinogluInPrep} can be expressed as
\begin{equation}\label{eqEffectiveCoupling}
  \tilde{g}(t) = \frac{1}{\sqrt{2}}
  \frac{g\alpha}{\Delta(\Delta+\alpha)}
  \Omega(t).
\end{equation}
Within the second-order approximation valid for coupling $g$ and drive strength $\Omega$ much smaller than $|\Delta|$ and $|\Delta+\alpha|$, the strength of the effective coupling is limited to $|\tilde{g}(t)|\ll |\alpha|$. In typical transmon circuits with anharmonicities on the order of few hundred MHz this restricts $\tilde{g}/2\pi$ to values below approximately $10\,\mathrm{MHz}$ and, thus, the shortest time in which a photon can be generated using this scheme to roughly $\pi/\tilde{g}\gtrsim 50\,\mathrm{ns}$. To suppress decoherence during the emission process, stronger effective couplings and therefore shorter photon pulses may be achievable with more anharmonic qubits such as the fluxonium \cite{Manucharyan2009}. Alternatively, transmon geometries with longer coherence times \cite{Rigetti2012,Barends2013,Chang2013} may be employed.

In analogy to the optical domain experiment with atoms in \cite{Keller2004}, we use the tunable effective coupling to generate single photons by controllably transferring population of the $|f0\rangle$ state into $|g1\rangle$ which decays into $|g0\rangle$ by photon emission. Since the drive is off-resonant from the $|g0\rangle\to |e0\rangle$ transition, the system remains trapped in the ground state [Fig.~\ref{figLevelScheme}(b)], ensuring that only a single photon is emitted. By modulating the amplitude and the phase of the drive signal in time using sideband mixing, we control the temporal shape of the output field $a_{\mathrm{out}}(t) = \sqrt{\kappa}a(t)$ \cite{Gardiner1985} resulting in the emission of a single photon  state $|1\rangle = \int{\psi(t) a_{\mathrm{out}}^{\dagger}(t)|0\rangle\,\mathrm{d}t}$ characterized by its mode function $\psi(t)$.

For the initial characterization of the created single photon states $|1\rangle$ we prefer to analyze their superposition with vacuum $(|0\rangle + |1\rangle)/\sqrt{2}$. In contrast to a single photon Fock state $|1\rangle$ this superposition state has a non-zero average voltage proportional to the output field $\langle a_{\mathrm{out}}(t) \rangle$ \cite{daSilva2010} given by $\psi(t)/2$. In circuit QED this voltage is readily determined using standard heterodyne measurements which also capture the phase of the mode function $\psi(t)$.
In addition we characterize single photon Fock states $|1\rangle$ directly by measuring the emitted power proportional to $\langle a_{\mathrm{out}}^{\dagger}(t) a_{\mathrm{out}}(t) \rangle = |\psi(t)|^2$ and also higher order moments of the output field which allow us to extract their full density matrices \cite{Bozyigit2011,Eichler2012}.

\begin{figure}
\includegraphics[scale=0.57]{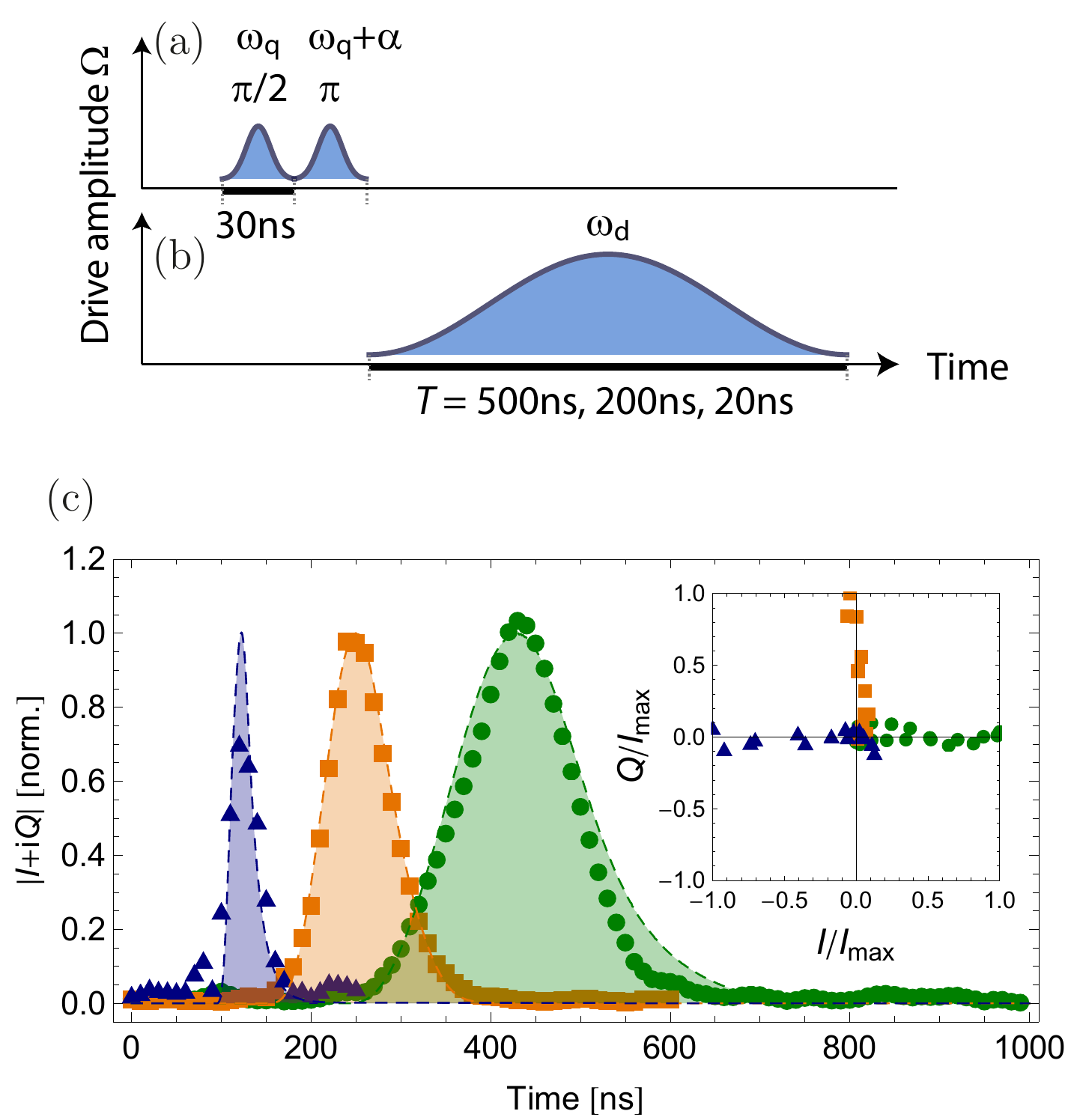}
\caption{(a) Amplitude of the drive signal used to initialize the transmon and (b) transfer the excitation into the resonator to generate a symmetric photon shape. (c) The normalized voltage amplitude $|I+\mathrm{i}Q|$ of shaped photons obtained for drive pulses with $T=20\,\mathrm{ns}$, $\Omega_0/2\pi = 680\,\mathrm{MHz}$ (blue triangles), $T=200\,\mathrm{ns}$, $\Omega_0/2\pi = 700\,\mathrm{MHz}$ (orange squares) and $T=500\,\mathrm{ns}$, $\Omega_0/2\pi = 600\,\mathrm{MHz}$ (green circles). The dashed lines show the simulated photon shapes scaled to normalize their peak values to unity. The measured traces are fitted to the simulation with only the time-shifts and scaling factors as fit parameters. The inset shows both voltage quadratures $I$ and $Q$ of the photon pulses.}
\label{figSymPhoton}
\end{figure}

In order to prepare the photon superposition state $(|0\rangle + |1\rangle)/\sqrt{2}$, we drive the $|g\rangle\to |e\rangle$ and $|e\rangle\to |f\rangle$ transitions sequentially [Fig.~\ref{figSymPhoton}(a)] to initialize the transmon in the state $(|g\rangle+|f\rangle)/\sqrt{2}$. This state is then mapped coherently onto $(|0\rangle + |1\rangle)/\sqrt{2}$ by driving the second-order process discussed above with a pulse [Fig.~\ref{figSymPhoton}(b)] of the form
\begin{equation}
\Omega(t) = \Omega_0 \sin^2(\pi t/T) \exp(\mathrm{i}\phi(t)).
\label{drivesignal}
\end{equation}
The pulse $\Omega(t)$ is parametrized by its maximum amplitude $\Omega_0$, its duration $T$ and its phase $\phi(t)$. We adjust the phase $\phi(t)$ to compensate for the amplitude-dependent Stark shift $\Delta_{f0g1}(t)$, keeping the phase of the photon waveform constant over the duration of the pulse (see Appendix~\ref{appendixCal}). We choose this $\sin^2$ shape of the drive pulse as a particularly simple continuous function with only two free parameters and -- unlike other candidates such as a Gaussian function -- with no need for an arbitrary cutoff.

To determine the waveform of the photon state, we measure both voltage quadratures $I(t)$ and $Q(t)$ of the output field emitted into the detection line in a heterodyne setup \cite{Bozyigit2011}.  The averaged value $\langle v(t)\rangle$ of the complex signal $v(t) = I(t)+\mathrm{i}Q(t)$ is then proportional to the expectation value $\langle a_{\mathrm{out}}(t) \rangle = \psi(t)/2$. To detect the weak single-photon signal, a broadband low-noise high electron mobility transistor (HEMT) amplifier is used. For measurements requiring only narrow bandwidth not exceeding approximately $10\,\mathrm{MHz}$, we also employ a phase-preserving Josephson parametric amplifier \cite{Bergeal2010,Eichler2014b} to operate at higher signal-to-noise ratio \cite{Eichler2012}. A boxcar filter is applied to the digitized signal to reduce noise and to filter out the pump tone of the parametric amplifier as well as to compensate the DC offset of the A/D converter.

We perform measurements of the photon waveform for different values of $T$ and $\Omega_0$ to study the dependence of the photon shape on these parameters (see Appendix \ref{appendixCal} and Fig.~\ref{figShapingCalib}(a,b)) and to maximize its symmetry. We quantify the photon symmetry using a parameter $s$ given by the scalar product of the waveform with its time-inverse (Eq.~(\ref{eqSymParam})). 
By choosing $T$ and $\Omega_0$ resulting in high values of the symmetry parameter $s$, we are able to prepare in good approximation symmetric photons of controlled length, as shown in Fig.~\ref{figSymPhoton}(c). Performing qubit tomography measurements after the photon emission process shows that for the two longer photon pulses ($T=200\,\mathrm{ns}$, $T=500\,\mathrm{ns}$) the initial $|f0\rangle$ state is nearly emptied by the drive pulse, having a residual population on the order of $1-2\%$. On comparable time scales the reduction of the population due to relaxation is negligible. The peak drive strengths for these pulses are $\Omega_0/2\pi = 700\,\mathrm{MHz}$ and $\Omega_0/2\pi = 600\,\mathrm{MHz}$, respectively. The corresponding peak amplitudes of the effective coupling $\tilde{g}/2\pi$ given by Eq.~(\ref{eqEffectiveCoupling}) are $5.2\,\mathrm{MHz}$ and $4.4\,\mathrm{MHz}$, consistent with numerical diagonalization of the Hamiltonian which yields values of $\tilde{g}/2\pi = 5.5\,\mathrm{MHz}$ and $4.6\,\mathrm{MHz}$. The symmetry parameter $s$ reaches a value of $0.98$ for both longer pulses. This high symmetry can only be obtained for pulses much longer than the cavity rise time $1/\kappa\approx 7\,\mathrm{ns}$. For comparison, we show a short pulse with $T = 20\,\mathrm{ns}$. This photon pulse is not long enough for a complete population transfer from $|f0\rangle$ to $|g1\rangle$ with the drive pulse amplitudes used which leads to a reduced emission efficiency. Its symmetry parameter $s = 0.92$ also does not reach the high values obtained for the longer pulses.

To prepare a symmetric photon state which can be easily reabsorbed by a quantum node, it is important that not only its amplitude but also its phase is symmetric in time. We achieve this using the AC Stark shift calibration procedure described in Appendix~\ref{appendixCal}. By choosing the time-dependent phase $\phi(t)$ of the drive pulse given by Eq.~(\ref{drivesignal}) such that $\dot{\phi}(t) = -\Delta_{f0g1}(t)$, we make the phase of the photon pulse constant in time, as illustrated in the inset of Fig.~\ref{figSymPhoton}(c), noting that the ratio of the two signal quadratures is in good approximation time-independent for each of the three pulses.

The system dynamics including imperfections of the second-order transition and decoherence are modelled by solving the master equation for the full Jaynes-Cummings Hamiltonian including three resonator and six transmon levels and evaluating the output field $a_{\mathrm{out}}(t)$ to obtain the simulation data in Fig.~\ref{figSymPhoton}(c).

To demonstrate the single photon nature of the emitted field, we measure its moments using the propagating field tomography method described in detail in Refs.~\cite{Eichler2012,Menzel2010}. We first determine the mode function $\psi(t)$ of the photon pulse which is proportional to the observed averaged coherent signal $\langle v(t)\rangle$. Then in each realization of the experiment, the measured single-shot voltage $v(t)$ is processed by a digital Chebyshev filter with a shape approximately matched to $\psi(t)$ and the result $V$ is recorded in a 2D histogram. The observable $V$ can be described by an operator $A+h^{\dagger}$ where $A$ is the temporal field mode $A = \int{\psi^{*}(t) a_{\mathrm{out}}(t)\,\mathrm{d}t}$ and $h$ is a noise mode, assumed to be in a thermal state which is characterized by the effective noise temperature of the amplification chain \cite{Eichler2012}. The moments $\langle (V^{*})^{m} V^{n}\rangle$ extracted from the recorded histogram and the noise moments $\langle (h^{\dagger})^{k} h^{l}\rangle$ determined in an equivalent measurement with only vacuum at the input of the detection chain are used to calculate the field moments $\langle (A^{\dagger})^{i} A^{j}\rangle$ \cite{Eichler2012}.

\begin{figure}[!b]
\includegraphics[width=8.5cm]{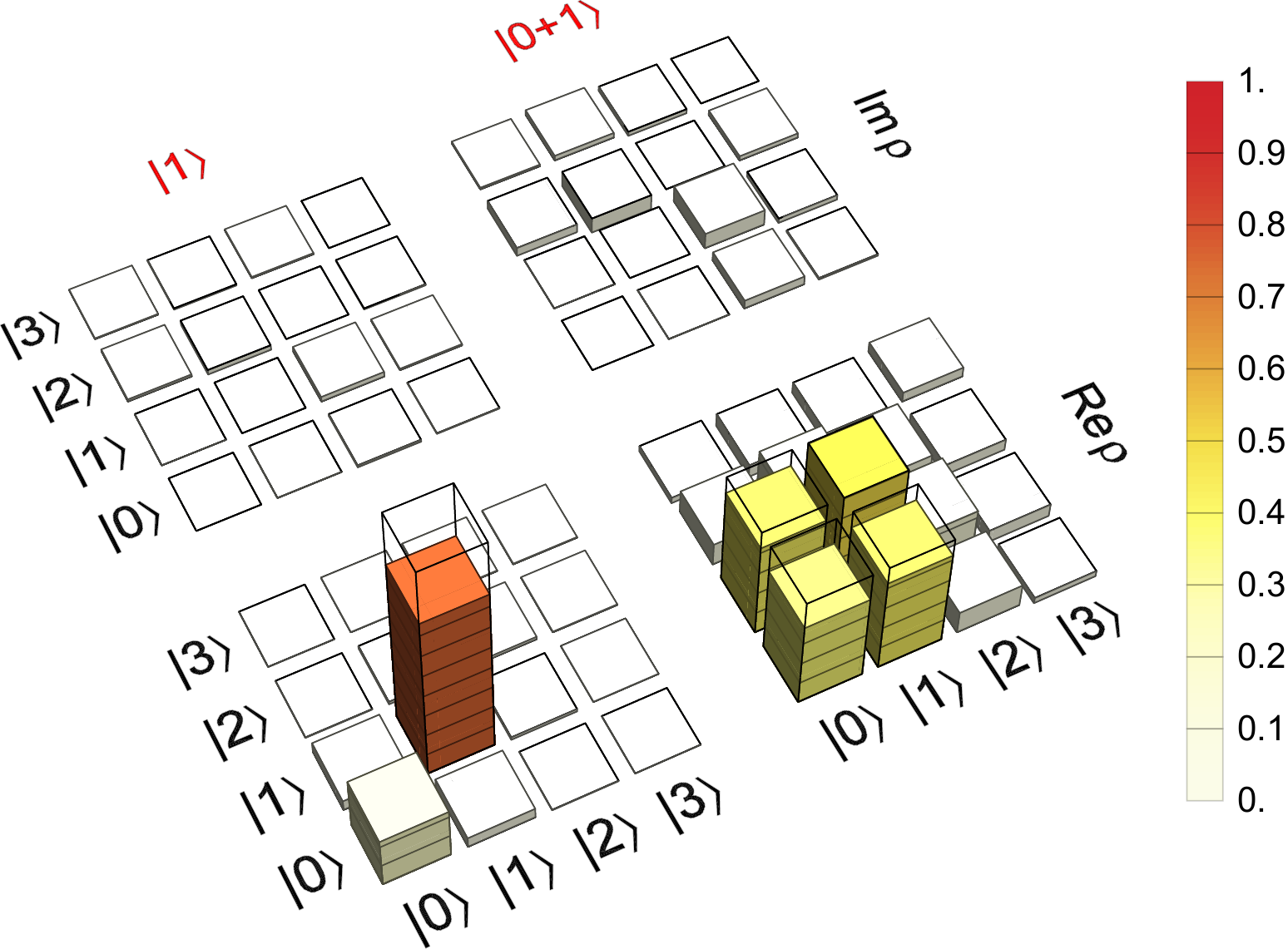}
\caption{The real and imaginary part of the measured density matrices $\rho$ of the symmetric temporal photon mode with fidelities $F=76\%$ and $F=86\%$ to the respective ideal states $|1\rangle$ and $(|0\rangle + |1\rangle) / \sqrt{2}$. The wire frames show the single-photon Fock state $|1\rangle$ (all density matrix elements equal to $1$ or $0$) and the ideal superposition state $(|0\rangle+|1\rangle)/\sqrt{2}$ (all density matrix elements equal to $1/2$ or $0$). The ticks in the bars represent increments of $0.1$.}
\label{figDensityMatrix}
\end{figure}

For both the symmetrically shaped photon superposition state $(|0\rangle+|1\rangle)/\sqrt{2}$ and the single photon Fock state $|1\rangle$ described above, we find the normalized fourth order moments $g^{(2)}(0) = \langle A^{\dagger}A^{\dagger}AA\rangle / \langle A^{\dagger}A\rangle^2$ of $0.03\pm 0.07$ and $0.06\pm 0.02$, respectively, which lie well below the classical limit of $g^{(2)}(0) = 1$ expected for coherent states, showing a high degree of antibunching. The density matrices $\rho$ with fidelities $F=86\%$ and $76\%$ of the respective photon states shown in Fig.~\ref{figDensityMatrix} are extracted from the measured moments \cite{Eichler2012} by employing a maximum likelihood algorithm. From the numerical simulation of the emission process, we find the normalization conditions $\langle A^{\dagger}A\rangle = 0.39$ and $0.79$. This is lower than the values of $\langle A^{\dagger}A\rangle = 1/2$ and $1$ expected for the ideal states $(|0\rangle+|1\rangle)/\sqrt{2}$ and $|1\rangle$ due to the reduced photon emission efficiency of $(79\pm 1)\%$ limited by the finite lifetime $T_1^f=(550\pm 5)\,\mathrm{ns}$ of the $|f\rangle$ state. This accounts for the deviation of the diagonal elements of $\rho$ from the theoretical values while the off-diagonal elements are reduced due to the loss of coherence between the qubit states $|g\rangle$ and $|f\rangle$ on a time scale of $T_2^{gf}= (580\pm 30)\,\mathrm{ns}$. This estimate of the emission efficiency assumes perfect initialization of the transmon in the $|f\rangle$ state in order to evaluate the shaping process separately from the preparation procedure. The total efficiency including a realistic initial state preparation is approximately $76\%$, that is, about $6\%$ lower due to relaxation during the initialization pulses and thermal population of the excited states (see Appendix \ref{appendixSys}).


\begin{figure}
\includegraphics[width=8.5cm]{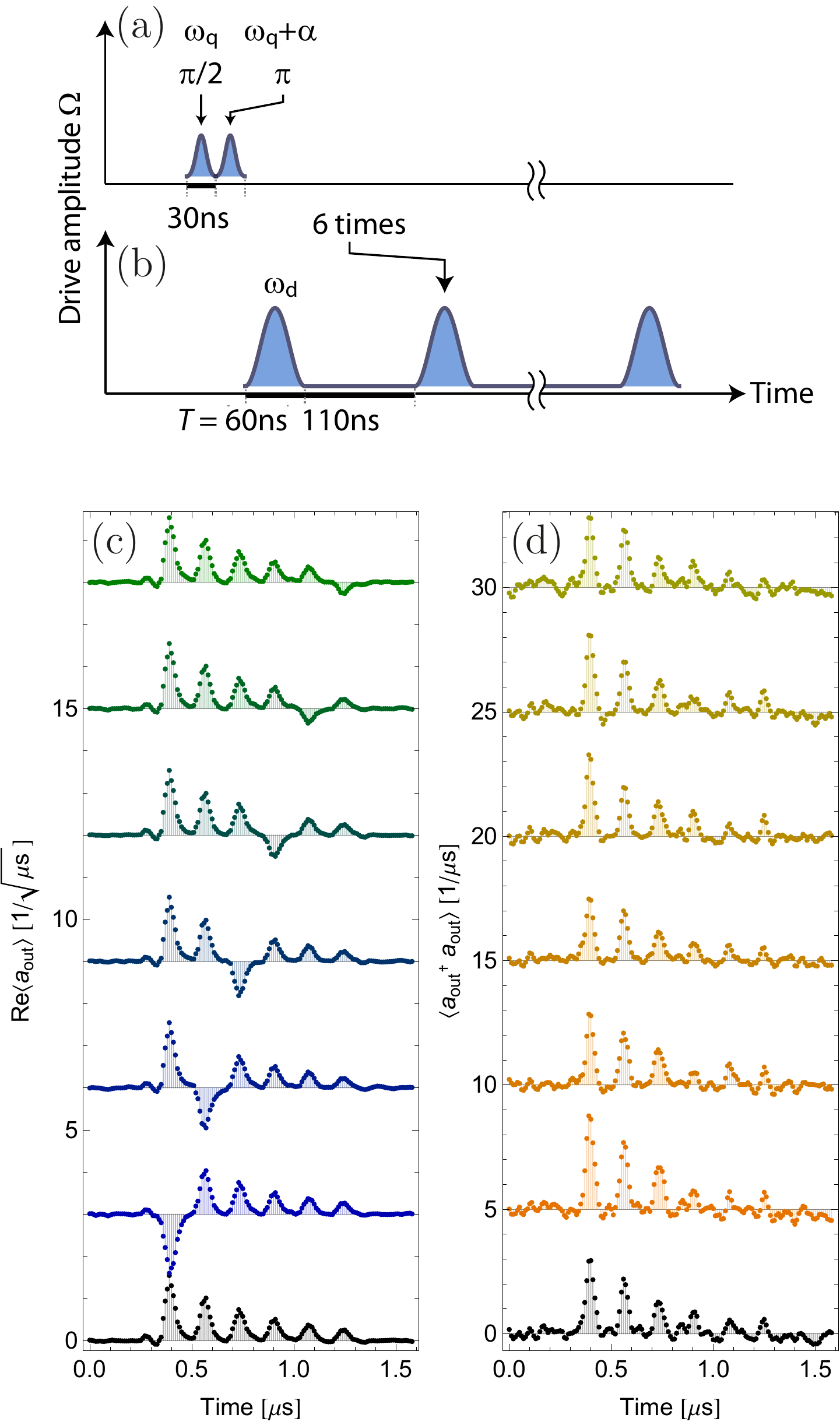}
\caption{(a) Amplitude of the drive signal used to initialize the transmon and (b) transfer the excitation into the resonator to generate a multi-peaked photon shape. (c) The real part of the measured output field $a_{\mathrm{out}}$ as a function of time for six-peak photons with phases of each of the peaks flipped by $\pi$ one by one. The bottom-most trace is a reference with all peaks having the same phase. (d) Averaged emitted power as a function of time for the same photon pulses as in (c).}
\label{figDoublePeakPhoton}
\end{figure}

To demonstrate the rapid amplitude and phase modulation capability of our all-microwave photon shaping scheme we prepare six-peaked single photon pulses similar to the double-peaked pulses demonstrated with optical frequency photons \cite{Keller2004}. For this purpose the transmon is again prepared in the state $(|g\rangle+|f\rangle)/\sqrt{2}$ [Fig.~\ref{figDoublePeakPhoton}(a)]. The subsequent photon shaping signal consists of a train of six identical $\sin^2$ pulses of amplitude $\Omega_0/2\pi \approx 350\,\mathrm{MHz}$ and length $T=60\,\mathrm{ns}$ separated by $170\,\mathrm{ns}$ [Fig.~\ref{figDoublePeakPhoton}(b)]. These parameters are adjusted to make the overlap between the photon peaks small while keeping the overall duration of the pulse train short to minimize decoherence.

We show that the phases of the individual peaks in the photon waveform can be controlled independently.  As an example we changed the phase of any one of the subsequent photon peaks by $\pi$ by adjusting the phase of the corresponding drive pulses leading to a change of sign in the detected voltage, Fig.~\ref{figDoublePeakPhoton}(c). This phase control can be achieved while keeping the emitted power unchanged, as illustrated in Fig.~\ref{figDoublePeakPhoton}(d). The power is measured as a function of time by squaring and subsequent averaging of the digitized and filtered voltages. The noise power is then subtracted in post-processing. 

In conclusion, we have presented a photon shaping technique relying fully on a phase- and amplitude-controlled microwave drive acting on three transmon levels. We have shown that this method can be used to generate single photon pulses of symmetric shapes with a controllable amplitude and phase. We have prepared a multi-peaked photon with individually tunable phases of all the peaks. Such photon states may be used to encode information in the rich multi-dimensional time-bin Hilbert space of a travelling photon \cite{Bechmann-Pasquinucci2000}. The shaping method can be refined by using simulations and optimization techniques to find drive pulses needed to generate desired photon shapes. With the time-reversed scheme applied to a second distant qubit the emitted photon can in principle be reabsorbed to map its quantum state onto a qubit \cite{Cirac1997}. The simple nature of this scheme with respect to the required control elements makes it also a prime candidate for use in 3D circuit QED architectures \cite{Paik2011}.

This work was supported by the European Research Council (ERC) through a Starting grant, by the Swiss National Science Foundation through the National Center of Competence in Research ``Quantum Science and Technology'' and by ETH Zurich.


\appendix

\section{System parameters}\label{appendixSys}

In this appendix we provide supplementary information about parameters of the sample, measurement setup and methods.

\begin{figure}[b]
\includegraphics[width=8.5cm]{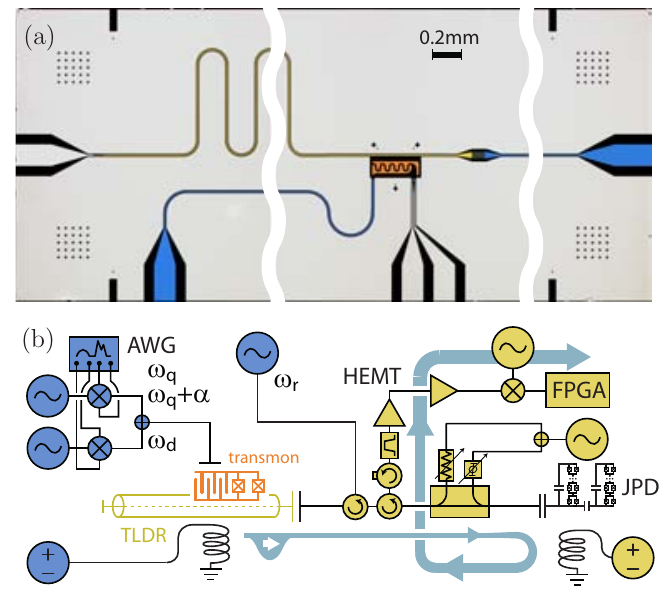}
\caption{(a) A false colour microscope image of the chip showing the transmission line resonator coupled to the transmon circuit (rectangular structure in the middle) and the microwave control lines: the resonator input/output port, the transmon charge line (blue) and the unused flux line (grey). (b) Schematic of the measurements setup (see Appendix \ref{appendixSys}). The control electronics is shown in blue, the measurement chain in yellow and the transmon circuit coupled to the transmission line resonator (TLR) in orange.}
\label{figSetup}
\end{figure}

We pattern the transmission line resonator using photolithography on a niobium-coated sapphire wafer [Fig.~\ref{figSetup}(a)]. The transmon circuit is fabricated by e-beam lithography and shadow evaporation of aluminium. We measure the charging energy $E_C/h\approx 406\,\mathrm{MHz}$ and maximum Josephson energy $E_J^{\mathrm{max}}/h\approx 25\,\mathrm{GHz}$ of the device. To tune the transition frequency of the transmon, we apply a magnetic flux bias using a superconducting coil and to measure its quantum state, we employ dispersive readout \cite{Bianchetti2009} by probing the reflection coefficient of the resonator [Fig.~\ref{figSetup}(b)].

The microwave pulses used to initialize the transmon in the $|f0\rangle$ state and to transfer its excitation into the resonator are generated by single sideband mixing using In-phase/Quadrature channels synthesized by an arbitrary waveform generator (AWG).

The output of the transmission line resonator is routed to a Josephson parametric dimer amplifier of the type described in \cite{Eichler2014b} via two circulators -- the first serving to separate the incoming drive signal from the output of the resonator, the second to separate the input and output of the parametric amplifier -- and a directional coupler used to pump the amplifier [Fig.~\ref{figSetup}(b)]. The pump tone is split and applied to two inputs of the directional coupler with relative phase and amplitude chosen to ensure destructive interference of the pump tone in the amplified signal. We operate the parametric amplifier in the phase preserving mode with a pump detuned from the signal by $240\,\mathrm{MHz}$. Its gain is adjusted to approximately $20\,\mathrm{dB}$ with a bandwidth of $30\,\mathrm{MHz}$. For measurements which do not require the added gain, the parametric amplifier is not pumped and the signal is reflected with a gain of unity. The signal is then filtered by a 4-8 GHz bandpass filter and amplified by a cryogenic HEMT amplifier with a gain of $35\,\mathrm{dB}$. At room temperature the signal is further amplified by $60\,\mathrm{dB}$ and downconverted with a local oscillator detuned by $25\,\mathrm{MHz}$ from the resonator frequency. The resulting intermediate frequency signal is digitized by an A/D converter at a rate of $100\,\mathrm{MS}/\mathrm{s}$. The two quadrature components of the IF signal are then determined by digital downconversion and low-pass filtering and further averaged. Alternatively, their values are integrated and the result recorded in a histogram. The efficiency of the detection chain is characterized by comparing the measured noise floor with the measured power of a single photon pulse, resulting in a noise number of $N_{0}\approx 10$ which corresponds to a detection efficiency of approximately $9\%$. This efficiency is limited predominantly by the losses between the sample and the parametric amplifier, the noise added by the detection chain and imperfections in mode matching due to the used Chebyshev digital filter being only an approximation of the true photon shape.

The state preparation pulses driving the qubits from $|g\rangle$ to $|e\rangle$ and from $|e\rangle$ to $|f\rangle$ are Gaussian with a standard deviation $\sigma = 5\,\mathrm{ns}$ truncated to a finite length of $6\sigma$. We calibrate their amplitudes by Rabi oscillation measurements and their frequencies using Ramsey interferometry. The latter measurement is also used to extract the dephasing times of the transmon $T_2^{ge}=(1640\pm 50)\,\mathrm{ns}$, $T_2^{ef}=(557\pm 8)\,\mathrm{ns}$ and $T_2^{gf}=(580\pm 30)\,\mathrm{ns}$. The relaxation times $T_1^{e}=(2000\pm 200)\,\mathrm{ns}$ and $T_1^{f}=(550\pm 5)\,\mathrm{ns}$ are determined by time-resolved measurements of the excited state population using three-level quantum state tomography \cite{Bianchetti2010}.

The steady state thermal population of the first excited state of the transmon is measured to be approximately $13\%$, significantly higher than the theoretical equilibrium value $0.2\%$ corresponding to the physical base temperature of $50\,\mathrm{mK}$. The source of the excess thermal population is likely due to undesired elevated temperatures of the still and $100\,\rm{mK}$ stages of the employed cryostat. To prepare the transmon in its ground state, we swap the thermal population of the first excited state $|e\rangle$ into $|f\rangle$ and make use of the $|f0\rangle\to|g1\rangle$ transition to further transfer it into the resonator. As the resonator relaxation rate $\kappa$ is much higher than the qubit decay rate, this results in a decrease of the total system energy. By repeating the cooling step several times, we reach $|e\rangle$ state thermal population of approximately $3\%$. This also indicates that the mean thermal excitation of the resonator is significantly lower than that of the qubit.

\section{Calibration of the photon-shaping drive pulse}\label{appendixCal}
This appendix provides explanation of the methods used to calibrate the photon shaping drive pulses.

\begin{figure*}
\includegraphics[width=18.3cm]{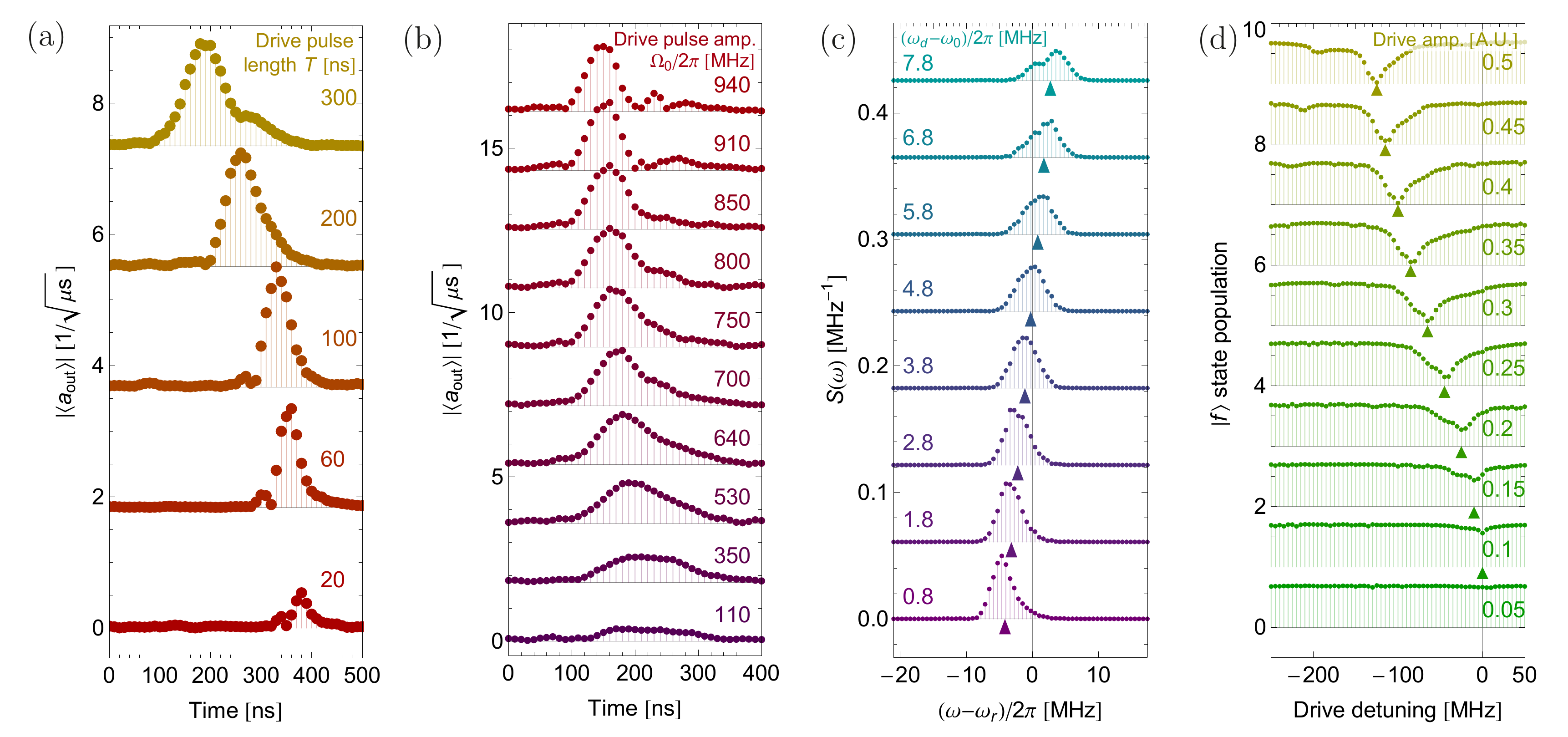}
\caption{(a) The measured voltage waveform of the photon superposition state $(|0\rangle+|1\rangle)/\sqrt{2}$ for the indicated length of the drive pulse $T$ at a fixed pulse amplitude $\Omega_0/2\pi = 420\,\mathrm{MHz}$ and (b) for the indicated amplitude of the pulse $\Omega_0$ at a fixed pulse length $T=300\,\mathrm{ns}$. (c) Fourier transform of the measured photon pulse waveform obtained for $\Omega_0/2\pi = 600\,\mathrm{MHz}$ and $T = 500\,\mathrm{ns}$ in a reference frame rotating at the resonator frequency $\omega_r$. The different curves correspond to the indicated values of the detuning between the drive frequency $\omega_d$ and the expected zero-amplitude limit of the $|f0\rangle\to |g1\rangle$ transition frequency $\omega_0 = 2\omega_q+\alpha-\omega_r$. The triangle marks the center frequency of the peak. (d) Population of the $|f\rangle$ state after a $100\,\mathrm{ns}$ square pulse driving the $|f0\rangle\to|g1\rangle$ transition plotted versus detuning of the pulse from $\omega_0$ for the indicated pulse amplitudes (shown here in units of the full scale output amplitude of the AWG). The triangles mark the minima of the curves.}
\label{figShapingCalib}
\end{figure*}

To determine the optimal values of $\Omega_0$ and $T$ in the drive pulse (\ref{drivesignal}) for generating a symmetric photon shape, we measure the heterodyne voltage of the shaped superposition state $(|0\rangle+|1\rangle)/\sqrt{2}$ for a range of $T$ between $60\,\mathrm{ns}$ and $500\,\mathrm{ns}$ and $\Omega_0/2\pi$ between $0\,\mathrm{MHz}$ and $1000\,\mathrm{MHz}$. A set of the measured photon pulse waveforms chosen to illustrate the dependence of the shape on $\Omega_0$ and $T$ is shown in Fig.~\ref{figShapingCalib}(a,b).

The observed effect of the drive pulse length on the length of the photon pulse and the efficiency of its emission, which is reflected in the amplitude of the detected voltage, is illustrated in Fig.~\ref{figShapingCalib}(a). Shorter drive pulses lead to shorter photon waveforms (curves at the bottom of the plot) and vice versa. 

The influence of the drive pulse amplitude is shown in Fig.~\ref{figShapingCalib}(b). Photon shapes generated with stronger drive pulses are shorter and display signatures of Rabi oscillations (the smaller side-peaks in the two curves at the top of the plot). On the other hand, weak pulses result in an incomplete population transfer and therefore reduced efficiency of the emission process, as illustrated by the reduced amplitude of the bottom-most waveform.

In order to quantify the symmetry of the measured waveform, we calculate the overlap $s$ of the averaged signal $\langle v(t)\rangle = \langle I(t)+\mathrm{i}Q(t)\rangle$ with the time-reversed copy of itself
\begin{equation}\label{eqSymParam}
  s = \max_{t_0}
  \frac{|\int{\langle v(2t_0-t)\rangle^{*} \langle v(t)\rangle}\,\mathrm{d}t|}
  {\int{|\langle v(t)\rangle|^2}\,\mathrm{d}t}.
\end{equation}
We then numerically maximize $s$ with respect to the amplitude $\Omega_0$ and length $T$ of the drive pulse.
We find values of $s$ close to one for various combinations of these parameters, e.g. $s = 0.98$ for $\Omega_0/2\pi = 700\,\mathrm{MHz}$, $T = 200\,\mathrm{ns}$ and $s = 0.99$ for $\Omega_0/2\pi = 600\,\mathrm{MHz}$, $T = 500\,\mathrm{ns}$, whose corresponding photon waveforms are also shown in Fig.~\ref{figSymPhoton}(c).

The maximization of the symmetry parameter also helps us fine-tune the frequency of the drive pulse to resonance with the $|f0\rangle\to|g1\rangle$ transition. Since a constant detuning leads to a linear phase drift in the emitted photon waveform, the symmetry value $s$ is reduced when the drive pulse is off-resonant. Alternatively, we can fine-tune the drive frequency using direct measurements of the photon frequency as shown in Fig.~\ref{figShapingCalib}(c). Here we plot the Fourier transform of the photon waveform and observe that its peak shifts as the drive frequency $\omega_d$ is tuned across the $|f0\rangle\to|g1\rangle$ transition frequency $\omega_0 = 2\omega_q + \alpha - \omega_r$ expected in the zero-amplitude limit. We then adjust the drive frequency to eliminate the detuning of the photon. The results of these two methods for calibrating the drive frequency are consistent with each other.

The conversion between the output amplitude of the drive signal set at the AWG and the drive parameter $\Omega(t)$ is obtained using measurements of the AC Stark shift presented in Fig.~\ref{figShapingCalib}(d). We prepare the transmon in the $|f\rangle$ state, apply pulses with different frequencies close to $\omega_0$ and extract the drive detuning at which the $|f\rangle$ state is maximally depleted after the pulse. This detuning determines the Stark shift. By repeating the measurement for different amplitudes of the drive pulse and comparing the obtained Stark shifts with the results of numerical diagonalization of the transmon-resonator Hamiltonian, we can match each AWG output amplitude with the corresponding value of $\Omega_0$.

\end{document}